\documentclass[preprint,aps,fleqn]{revtex4}%
\usepackage{graphicx}
\usepackage{psfrag}
\usepackage{amsmath}
\usepackage{subfigure}

\begin{document}
\def\be{\begin{eqnarray}}
\def\ee{\end{eqnarray}}
\newcommand{\nn}{\nonumber}
\def\mpcomm#1{\nextline\strut\kern-6em{\tt MP COMMENT => \ #1}\nextline}
\def\nextline{\hfill\break}
\newcommand{\rf}[1]{(\ref{#1})}
\newcommand{\beq}{\begin{equation}}
\newcommand{\eeq}{\end{equation}}
\newcommand{\bea}{\begin{eqnarray}}
\newcommand{\eea}{\end{eqnarray}}
\newcommand{\pint}{-\hspace{-11pt}\int_{-\infty}^\infty }
\newcommand{\Pint}{-\hspace{-13pt}\int_{-\infty}^\infty }
\newcommand{\nint}{\int_{\infty}^{\infty}}
\renewcommand{\vec}[1]{\boldsymbol{#1}}

\title{Heavy Exotic Molecules}
\author{Yizhuang Liu}
\email{yizhuang.liu@stonybrook.edu}
\author{Ismail Zahed}
\email{ismail.zahed@stonybrook.edu}
\affiliation{Department of Physics and Astronomy\\
State University of New York, Stony Brook, NY 11794-3800}

{\begin{abstract}
We briefly review the formation of pion-mediated heavy-light  exotic molecules with both charm and bottom,
under the general strictures of chiral and heavy quark symmetries.  
The charm isosinglet exotic molecules with  $J^{PC}=1^{++}$ binds,
which we identify as the reported neutral $X(3872)$. The bottom isotriplet exotic with $J^{PC}=1^{+-}$ binds, 
and is identified as a mixed state of the reported charged exotics $Z^+_b(10610)$ and $Z^+_b(10650)$. 
The bound bottom isosinglet molecule with $J^{PC}=1^{++}$ is a possible neutral $X_b(10532)$ to be observed.
\vskip 10cm
\noindent {To appear in {\it Nuclei, Quarks, Stars} -memorial dedicated to Gerry Brown,\\
\noindent  Eds. J.W. Holt, T.T.S. Kuo, K.K. Phua, M. Rho
and I. Zahed, World Scientific (2016).}
\end{abstract}
}


 \maketitle

\section{Introduction}
\label{Intro_sec1}

A decade ago, the BaBar collaboration~\cite{BABAR}  and the CLEOII collaboration~\cite{CLEOII} 
have reported narrow peaks in the $D_s^+\pi^0$ (2317) and the $D_s^{*+}\pi^0$ (2460) in support
of predictions from chiral  and heavy-quark symmetry~\cite{MACIEK,BARDEEN,ISGUR}. The 
heavy-light multiplet  $(0^-, 1^-)=(D,D^*)$ has a chiral partner $(0^+, 1^+)=(\tilde D,{\tilde D}^*)$ that is
about  one consituent mass heavier~\cite{MACIEK,BARDEEN}. More recently, 
the Belle collaboration~\cite{BELLE} and the BESIII collaboration~\cite{BESIII} have reported new 
multiquark exotics, outside the standard quark model classification. A key source for these exotics 
is $\Upsilon(10860)$  and its closeness to the  $B\bar B^*\pi$ (10744) and $B^*\bar B^*\pi$ (10790) 
tresholds.  The smallness of the  phase space for the pion decay of $\Upsilon(10860)$ suggests that
the decay process is slow, involving a molecular configuration on the way out. 
Heavy exotic molecules have been reported,  such as the neutral $X(3872)$ and the charged 
$Z_c(3900)^\pm$ and $Z_b(10610)^\pm$. More of these exotics are expected to be unravelled by 
the DO collaboration at Fermilab~\cite{DO},  and the  LHCb collaboration at Cern~\cite{LHCb}.

Theoretical arguments have predicted the possibility of molecular bound states involving heavy-light
charm and bottom mesons through pion exchang~\cite{MOLECULES,THORSSON}. Since, 
a number of molecular estimates were made by many~\cite{THORSSON,KARLINER,OTHERS,OTHERSX,OTHERSZ,OTHERSXX}.
Non-molecular heavy exotics were also propodsed using constituent quark models~\cite{MANOHAR}, 
heavy solitonic baryons~\cite{RISKA,MACIEK2}, instantons~\cite{MACIEK3} and QCD sum rules~\cite{SUHONG}.

In this contribution, we briefly review our recent analysis of the molecular configurations with heavy-light
charm and bottom mesons and their chiral partners, under the general constraints of chiral and
heavy-quark symmetry~\cite{LIUHEAVY}. In section 2, we outline the heavy-light effective action 
in leading order involving the $(0^\pm, 1^\pm)$ multiplets, and formulate the non-relativistic bound
state problem in the $J=1$ channel. The results
are summarized in section 3. Our conclusions are in section 4.

\section{Molecules}

The leading part of  the heavy-light Lagrangian for the charmed multiplet $(0^-, 1^-)$ with pions reads~\cite{MACIEK,ISGUR}

\bea
\label{L-}
{\cal L}\approx &&+2i\left(\bar D \partial_0D+\vec{\bar D}\cdot \partial_0\vec{D}\right)
-\Delta m_{D}\bar DD-\Delta m_{\vec D}\vec{\bar D}\vec{D}\nonumber\\
&&+i\frac{g_H}{f_{\pi}}{\rm Tr}\partial_i\pi \left(D_iD^{\dagger}-DD_i^{\dagger}+\epsilon_{ijk}D_kD_j^{\dagger}\right)
\eea
with $\Delta m_i=m_i-m_C$ of the order of a quark constituent mass. 
The leading part of  the heavy-light chiral doublers  Lagrangian for the charmed 
$(0^+, 1^+)$ multiplet with pions reads~\cite{MACIEK}

\bea
\label{L+}
{\cal \tilde L}\approx &&+2i\left(\bar  {\tilde D} \partial_0{\tilde D}+\vec{\bar {\tilde D}}\cdot \partial_0\vec{\tilde D}\right)
-\Delta m_{\tilde D}\bar {\tilde D}\tilde D-\Delta m_{\vec {\tilde D}}\vec{\bar {\tilde D}}\vec{\tilde D}\nonumber\\
&&+i\frac{g_H}{f_{\pi}}{\rm Tr} \partial_i\pi \left(i(\tilde D_i{\tilde D}^{\dagger}+\tilde D\tilde D_i^{\dagger})
+\epsilon_{ijk}\tilde D_k{\tilde D}_j^{\dagger}\right)\nonumber\\
\eea
with again $\Delta m_{\tilde i}=m_{\tilde i}-m_C$ of the order of a quark constituent mass. 
The $(0^+,1^+)$ multiplet mixes with the $(0^-,1^-)$ by chiral symmetry~\cite{MACIEK,BARDEEN}

\be
\label{L+-}
{\delta \cal L} =\frac{g_{HG}}{f_{\pi}}{\rm Tr}\,\partial_0\pi \left({\tilde D}^{\dagger}_iD_i-i{\tilde D}^{\dagger}D+ {\rm c.c.}\right)
\ee

The molecular exotics of the type $D\bar D^*$, follows from (\ref{L-}) through one-pion exchange.
The non-relativistic character of the molecules yield naturally to a Hamiltonian description. Let
$D_{0\bar 0}(\vec{r})$ denote  the wave function of the molecular scalar,  and
$\bar Y_{0\bar i}(\vec r)$ and $Y_{i\bar 0}(\vec r)$ denote the wavefunctions of the molecular 
vectors, and $T_{i\bar j}(\vec{r})$ the wavefunction of the molecular tensors. Using (\ref{L-}-\ref{L+-})
for the 2-body interactions, we have

\bea
\label{HWF}
&&(VT)_{k\bar l}=C\epsilon_{kim}\epsilon_{\bar l\bar jn}\partial_{mn}V (r)T_{i\bar j}\nonumber\\
&&(VT)_{0\bar 0}=C\partial_{i\bar j}VT_{i\bar j}\nonumber\\
&&(V\bar Y)_{k\bar 0}=-C\partial_k\partial_{\bar j}V(r)\bar Y_{0\bar j}\nonumber\\
&&(VT)_{0\bar k}=C\epsilon_{\bar k\bar l j}\partial_i\partial_jV(r)T_{i\bar l}\nonumber\\
&&(VT)_{\bar 0 k}=C\epsilon_{ k l j}\partial_{\bar i}\partial_jV(r)T_{l\bar i}
\eea
with the isospin factor

\be
C=\vec{I_1}\cdot \vec{I_2}=\left(\left.\frac{1}{4}\right|_{I=1},-\left.\frac{3}{4}\right|_{I=0}\right)
\ee
Here $V(r)$ is the  regulated one-pion exchange using the standard monopole form factor by analogy with the
pion-nucleon form factor~\cite{BROWN}. It is defined with 
a core cutoff $\Lambda\gg m_\pi$~\cite{THORSSON,BROWN}

\be
\label{VRR}
&&V(r)=\left(\frac{g_H}{f_\pi}\right)^2
\frac 1{4\pi}\left(\frac{e^{-m_\pi r}}{r}-\frac{e^{-\Lambda r}}{r}-(\Lambda^2-m_\pi^2)\frac{e^{-\Lambda r}}{2\Lambda}\right)
\ee
Throughout,  we will use $g_H=0.6$~\cite{MACIEK,BARDEEN} and $\Lambda=1$ GeV. The choice
of $\Lambda$ is the major uncertainty in the molecular analysis. The one-pion exchange in 
(\ref{HWF}) induces a D-wave admixture much like in the deuteron as a proton-neutron molecule~\cite{BROWN}.
It is very different from one-gluon exchange in heavy quarkonia~\cite{MANOHAR}.

The pertinent projections onto the higher $J^{PC}$ channels of the molecular wavefunctions in (\ref{HWF})
require the use of both vector and higher tensor spherical harmonics~\cite{EDMONDS,THORN}. For $J=1$,
we will use the explicit forms quoted in~\cite{THORN} with the ${}^SL_J$ 
assignment completly specified. For the $(1^\mp,0^\mp)$ multiplets, there are 4 different $1^{PC}$ sectors

\bea
\label{J1LSJ}
1^{++}:&&T^{2,2}_{i\bar j}({}^5D_1),Y^{0+}_{i}({}^3S_1),Y^{2+}_{i}({}^3D_1)\nonumber\\
1^{--}:&&T^{0,1}_{i\bar j}({}^1P_1),T^{2,1}_{i\bar j}({}^5P_1),T^{2,3}_{i\bar j}({}^5F_1),\nonumber\\
&&Y^{1-}_{i}({}^3P_1),D^{1}({}^1P^\prime_1)\nonumber\\
1^{+-}:&&T^{1,0}_{i\bar j}({}^3S_1),T^{1,2}_{i\bar j}({}^3D_1),Y^{0-}_{i}({}^3S^\prime_1),Y^{2-}_{i}({}^3D^\prime_1)\nonumber\\
1^{-+}:&&T^{1,1}_{i\bar j}({}^3P_1),Y^{1+}_{i}({}^3P^\prime_1)
\eea
The normalized tensor harmonics are detailed in~\cite{LIUHEAVY,THORN}.

 \begin{figure}[h!]
 \begin{center}
  \includegraphics[width=6cm]{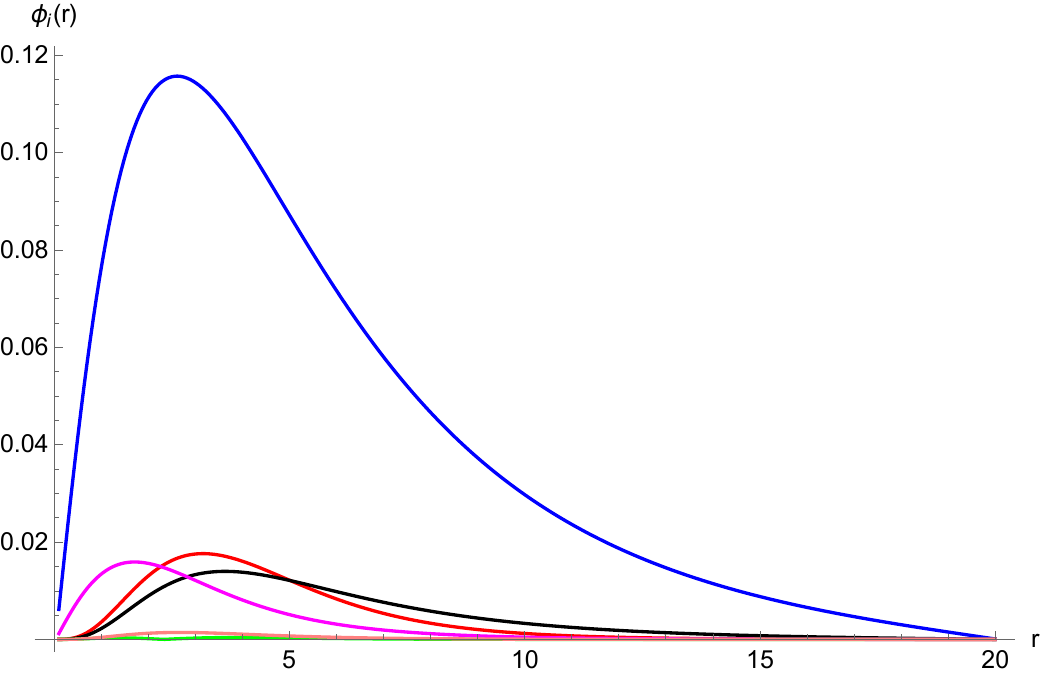}
   \includegraphics[width=6cm]{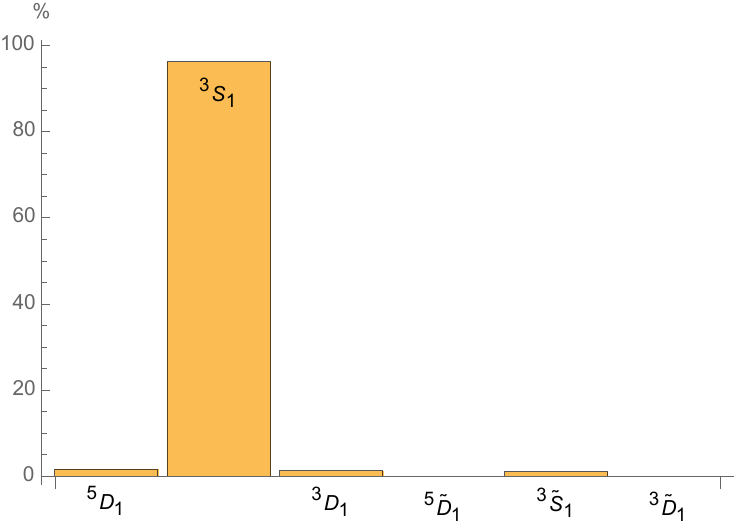}
   \caption{$J^{PC}=1^{++}$: radial wavefunctions  (upper plot) and percentage content (lower plot) 
  for the charm isosinglet exotic state ($C=-3/4$).}
    \label{fig_fcj1-1++}
  \end{center}
\end{figure}

\section{Results}

In Fig.~\ref{fig_fcj1-1++} we show the the radial components (upper part) and percentage content (lower part) 
of the bound  isosinglet charm wavefunction  with energy $E=3.867$ GeV, versus $r$ in units of 
$\Lambda =1$ GeV.  The intra-coupling between the $(0^-,1^-)$ and $(0^+,1^+)$ multiplet
causes the chiral partners  $\tilde D \bar{\tilde D}^*$ to unbind. The
${}^SL_J$ assignments referring to the $(0^-,1^-)$ multiplet, 
and the ${}^S\tilde L_J$ assignments referring to the $(0^+,1^+)$ multiplet, are those listed in (\ref{J1LSJ}).
The mixing results in a stronger binding in this channel wich is mostly an isosinglet ${}^1S_3$ contribution in the 
($1^-,0^-$) multiplet with almost no D-wave admixture. This molecular state carries $J^{PC}=1^{++}$.
It is chiefly  an isosinglet  $D\bar D^*$ molecule, which we identify as the reported exotic $X(3872)$.

 \begin{figure}[h!]
 \begin{center}
  \includegraphics[width=6cm]{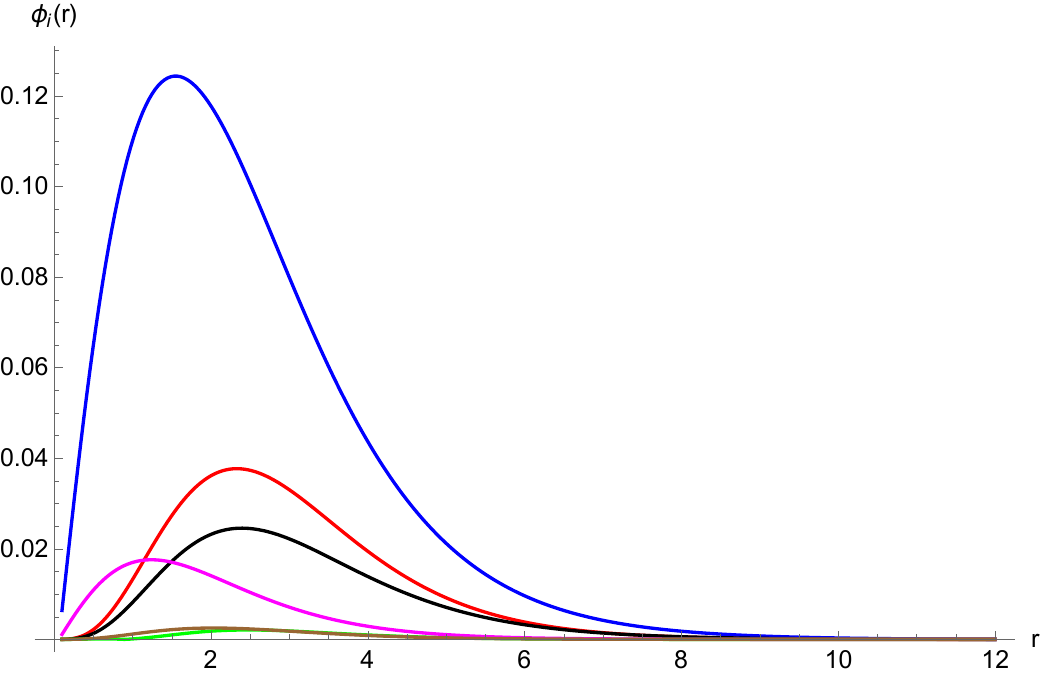}
   \includegraphics[width=6cm]{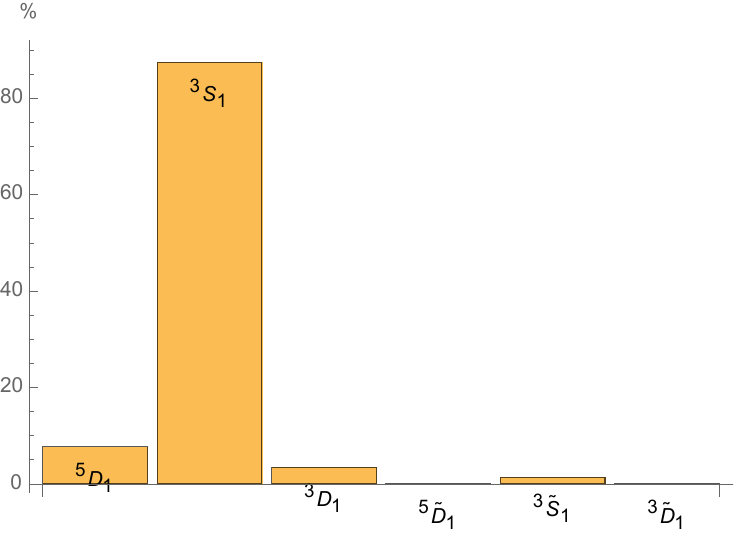}
   \caption{$J^{PC}=1^{++}$: radial wavefunctions  (upper plot) and percentage content (lower plot) 
  for the bottom isosinglet exotic state ($C=-3/4$).}
   \label{fig_fbj1-1++}
  \end{center}
\end{figure}

In Fig.~\ref{fig_fbj1-1++} we show the the radial components (upper part) and percentage content (lower part) 
of the bound  isosinglet bottom wavefunction  with energy $E=10.532$ GeV, versus $r$ in units of 
$\Lambda =1$ GeV. Again, the ${}^SL_J$ and ${}^S\tilde L_J$ assignments refer to the $(0^\pm, 1^\pm)$ multiplets respectively,
as defined in (\ref{J1LSJ}). The $1^{++}$ mixed bound state is mostly a $B\bar B^*$ (${}^3S_1$) molecule.
A comparison of Fig.~\ref{fig_fcj1-1++}  to  Fig.~\ref{fig_fbj1-1++} shows that this neutral bottom molecular state is the mirror
 analogue of the neutral charm molecular state or $X_b(10532)$, yet to be reported.

\begin{figure}[h!]
 \begin{center}
  \includegraphics[width=6cm]{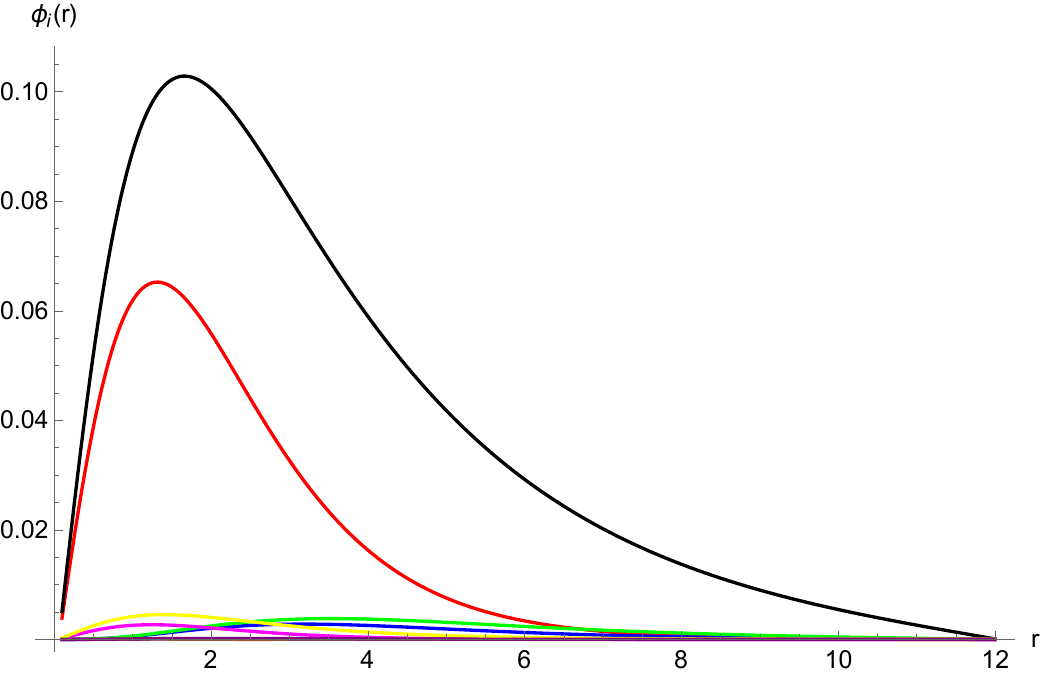}
   \includegraphics[width=6cm]{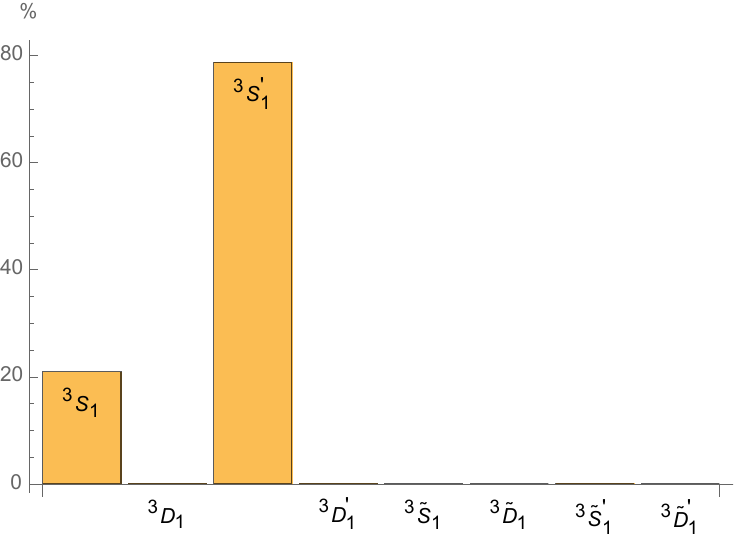}
  \caption{$J^{PC}=1^{+-}$: radial wavefunctions  (upper plot) and percentage content (lower plot) 
  for the bottom isotriplet  exotic state ($C=+1/4$).}
   \label{fig_fbj1-(1)1+-}
  \end{center}
\end{figure}

 In Fig.~\ref{fig_fbj1-(1)1+-}  we show the the radial components (upper part) and percentage content (lower part) 
of the bound  isosinglet bottom wavefunction  with energy  $E=10.592$ GeV, versus $r$ in units of 
$\Lambda =1$ GeV.  From the assignments given in (\ref{J1LSJ}),
it follows that $1^{+-}$ is a mixed isotriplet $B\bar B^*$ (${}^3S^\prime_1$) molecules,  with a small admixture of 
$B^*\bar B^*$ (${}^3S^\prime_1$) molecule. This molecule is an admixture of the reported states 
$Z^+_b(10610)$ and $Z^+_b(10650)$.

\section{Conclusions}
\label{SumCon}

We have briefly reported on the molecular states of doubly heavy mesons mediated by one-pion exchange for both the
chiral parteners $(0^\pm , 1^\pm)$ as a coupled channel problem, recently discussed in~\cite{LIUHEAVY}.
The  analysis complements and extends  those presented 
in~\cite{THORSSON,KARLINER,OTHERS,OTHERSX,OTHERSZ,OTHERSXX} by taking into 
account the strictures of chiral and heavy quark symmetry, and by retaining most coupled channels between the
$(0^-,1^-)$ multiplet and its chiral partner $(0^+,1^+)$. The key aspect of this coupling is to cause the
molecules in the $(0^-,1^-)$ multiplet to bind about twice more, and the molecules in the $(0^+,1^+)$ multiplet to unbind.
 The charm isosinglet exotic molecules with $J^{PC}=1^{++}$
is strictly bound for a pion-exchange cutoff $\Lambda=1 $ GeV. This state is identified with the reported isosinglet exotic 
X(3872) which in our case is mostly an isosinglet  $D\bar D^*$ molecule in the ${}^1S_0$ channel with no D-wave admixture. 
The attraction in the isotriplet channel with $J^{PC}=1^{+-}$ is too weak to bind the $D\bar D^*$ compound, suggesting that
the reported isotriplet $Z_C(3900)^\pm$ is at best a treshold enhancement.
The $Y(4260)$, $Y(4360)$ and $Y(4660)$ may point to the possibility of their constituents made of excited
$(D_1,D_2)$ heavy mesons and their chiral partners~\cite{MACIEK,EXCITED}.
The isotriplet bottom exotic molecule with $J^{PC}=1^{+-}$ 
which we have identified with the pair $Z^+_b(10610)$ and $Z^+_b(10650)$, which is a mixed state in our analysis.
The isosinglet bottom exotic molecule with $J^{PC}=1^{++}$ is a potential candidate for $X_b(10532)$, yet to be measured.

\section{Tribute to Gerry Brown}
\label{Tribs}

{\bf Ismail Zahed:}
Gerry  has been a mentor and a colleague  to me for the past three decades. He has taught many of us a great deal
of how to approach physics from the bottom line point of view and encouraged us  to pursue freely
our physics interests, always in the quest of tangible results. Most of us here at Stony Brook and many elsewhere, 
owe so much to Gerry scientific leadership and personal friendship. We hope to pass his legacy to our students 
and collaborators.

\section{Acknowledgements}

This work was supported by the U.S. Department of Energy under Contract No.
DE-FG-88ER40388.

%

\clearpage

%


\end{document}